# Morphological evolution of edge-hillocks on single crystal films having anisotropic drift-diffusion under the capillary and electromigration forces

Tarik Omer Ogurtani[a], Aytac Celik[b], and Emre Ersin Oren[c]

*Department of Metallurgical and Materials Engineering, Middle East Technical University, 06531, Ankara, Turkey*

**Abstract**

The morphological evolution of hillocks at the unpassivated sidewalls of the single crystal metallic thin films is investigated via computer simulations by using the free-moving boundary value problem. The effect of the drift-diffusion anisotropy on the development of surface topographical scenarios is fully explored under the action of electromigration and capillary forces, utilizing numerous combinations of the surface texture, the drift-diffusion anisotropy and the direction of the applied electric field. The present simulation studies yield very rich and technologically imported information, in regards to the critical texture of the single crystal thin film surfaces, and the intensity and the orientation of the applied electric field, as far as the device reliability is concerned.

[a] Corresponding author. Tel: +90 312 210 2512; Fax: +90 312 210 1267; e-mail: ogurtani@metu.edu.tr
url: http://www.csl.mete.metu.edu.tr

[b] E-mail: e104548@metu.edu.tr

[c] Current address: Materials Science and Engineering, Roberts Hall, Box: 352120, University of Washington, Seattle, WA, 98195; e-mail: eeoren@u.washington.edu

## 1. Introduction

The subject of capillary-driven morphological evolution at the surfaces and interfaces of condensed matters, especially under the action of applied force fields such as electrostatic and thermo-mechanical stress systems, has represented a challenging theoretical problem in materials science, without having exposed to any serious robust nonequilibrium thermodynamics treatments. The shortcomings of the classical thermodynamic, and its *ad hoc* applications that are hybridized by the phenomenological chemical kinetics in practice, were critically discussed and analyzed by Ogurtani and Oren [1]. This awkward situation is started to change very recently [2,3] because of the issue of surface evolution has found renewed interest over the past decade with the advancement in the nanotechnology [4]. The sub-microscopic feature of the electronic devices has been pushing the surfaces and interfaces into the front lines as the primary agents in the determination of the catastrophic failure of interconnects used in the microelectronic industry.

Theoretical studies of interconnect surfaces under the electromigration (EM) force have also revealed a variety of morphological scenarios. Krug and Dobbs [5] and Schimschak and Krug [6] showed that a crystal surface could be destabilized by an external electrostatic field in a material having anisotropic adatom surface diffusivity. Their linear instability analysis (LISA), like almost every analytical work published on the surfaces in the literature [7-9], strictly relies on the uniformly tilted surfaces and/or the small slope approximations, respectively. They also assumed that the field is almost constant along the surface. Nevertheless, it may be instructive in exploring the several aspects of the morphological instability of crystal surfaces and faceting transitions. In the later studies, Schimschak and Krug [10], Gungor and Maroudas [11], and Ogurtani and Oren [12] put more emphasis on the crucial role of the surface diffusion anisotropy and the crystalline texture in the development of the awkward morphological variations on the preexistent edge or internal voids causing catastrophic electrical break down. In all of these studies, the role of the boundary conditions on the dynamical behavior of the surfaces was under estimated. Gungor et al. [11] and Ogurtani and Oren [12,13] used rigid sidewalls attached to the edge void, which made the interconnect sidewalls partially inactive. Schimschak and Krug [10] used active edge surface properly, but they continue utilizing periodic boundary conditions in addition to the constant voltage application, which also hinder the fine details of the process, especially during the estimation of the effect of the current crowding. Only very recently, Ogurtani and Akyildiz [14] have considered the profound effects of the reflecting and/or free-moving boundary conditions on the EM induced grain boundary grooving (GBG) and cathode voiding, in their computer simulation studies. These computer experiments showed irrevocably that in the EM dominating regime the grain boundary (GB) voiding can be completely arrested by the applied current above the well defined threshold level.

Bradley [15] examined the effects of EM on the dynamics of corrugated interconnect-vapor interfaces using the multiple scale asymptotic analysis, by neglecting the capillary effects. This novel and very powerful technique was first introduced by Drazin and Johnson [16] to deduce the governing Korteweg-de Vries equation (KdV) equation for the irrotational 2-D motion of an incompressible and inviscid fluid. Bradley's [15] results are very interesting and predicting that the compressed and highly stretched solitary wave travels on the surface of a current carrying metallic thin film, having isotropic diffusivity, in the direction of the applied electrostatic field. The propagation velocity and the width of the solitons decrease with increasing amplitude. These observations are in contrast to the behavior of ordinary solitary waves in shallow waters [17,18], which are also governed by a KdV equation. The most interesting work on the EM induced edge stability in single-crystal metal lines was carried out by Mahadevan et al. [19] by employing a novel phase field (PF) technique to study this moving boundary problem numerically and assuming that the mobility of adatoms is anisotropic and having four-fold symmetry with $45^o$ tilt angle.

The most recent and extensive computer simulations are performed by Celik [20] on the finite size Gaussian shape edge voids chosen as initial configuration to induce Solitary waves without putting any physical and mathematical constrains on the model as advocated by Ogurtani and Oren [12]. These simulation experiments have irrevocably proved that even at the presence of strong diffusional anisotropy, the solitary waves (kinks or solitons and even a train of saw tooth waves) can be generated in the EM dominating regime, if one of the principal axes of the diffusivity dyadic has a special and irreducible orientation with respect to the applied electric field intensity vector.

*In situ* high voltage scanning electron microscope (HVSEM) examination of the accelerated EM tests performed by Marieb et al. [21], clearly showed the void nucleation at the metal-passivation interface of the Al-1% Si lines. These voids grew into the line in an inclined slit-like fashion, then begun to grow into a wedge-like shape. Several interesting phenomena were also observed that were not explained, such as, the liquid-like flow material during void coalescence and the dissolution of voids on the verge of the line breaching.

All these experimental and simulation observations are motivated us to program, execute and analyze robust computer simulations on the different combinations (192-variants) of the surface texture and tilt angle, the degree of anisotropy and EM wind intensity parameter, by a novel mathematical model supplemented by the tailor made computer coding.

## 2. Physical and Mathematical Modeling

The evolution kinematics of surfaces or interfacial layers (simply- or multiple-connected domains) may be described by the following well-posed moving boundary value problem in 2-D space (the generalized cylindrical surfaces in 3-D) for the ordinary points, in terms of normalized and scaled parameters and variables [1].

$$\bar{V}_{ord} = \frac{\partial}{\partial \bar{\ell}}\left[ D(\theta,\phi;m)\frac{\partial}{\partial \bar{\ell}}\left(\Delta \bar{g}_{vb} + \chi \bar{\vartheta} + \bar{\gamma}(\hat{\theta},\phi;m)\bar{\kappa}\right)\right] - \bar{M}_{vb}\left(\Delta \bar{g}_{vb} + \bar{\gamma}(\hat{\theta},\phi;m)\bar{\kappa}\right) \quad (1)$$

where, $\bar{\ell}$ is the curvilinear coordinate along the surface (arc length), $\bar{\kappa}$ is the local curvature and is taken to be positive for a convex void or a concave solid surface (troughs). $D(\theta,\phi;m) = 1 + A\cos^2[m(\theta-\phi)]$ and $\bar{\gamma}(\hat{\theta},\phi;m) = \left[\gamma(\hat{\theta},\phi;m) + \partial^2 \gamma(\hat{\theta},\phi;m)/\partial \hat{\theta}^2\right]$ are the angular parts of the anisotropic surface diffusion and surface specific Gibbs free energy (*the surface stiffness*), respectively. $\hat{\theta}$ and $\phi$ are, respectively, the angles between the line normal and the principal axis of the dyadic with respect to the applied electric field intensity vector in 2-D space; and similarly, $\theta$ is the angle between the tangent vector of the line profile and the electric field intensity vector. $A$ is the anisotropy constant, which may be a few orders of magnitude and $2m$ is the degree of fold of the symmetry (zone) axis. $\Delta \bar{g}_{vb} = (\bar{g}_v - \bar{g}_b)$ denotes the Gibbs free energy of transformation ($\Delta \bar{g}_{vb} < 0$ evaporation or void growth), which is the Gibbs free energy difference between the realistic void phase (vapor) and the bulk matrix, and it is normalized with respect to the minimum value of the specific surface Gibbs free energy of the interfacial layer denoted by $g_\sigma^o$. $\chi$ is the electron wind(EW) intensity parameter, $\bar{\vartheta}$ is the normalized electrostatic potential generated at the surface layer due to the applied electric field intensity.

In the above formula the surface drift-diffusion, which may be represented by an angular dependent post factor, $D(\theta, \phi; m)$ has been taken as anisotropic. On the contrary, for the time being the specific Gibbs free energy of the interfacial layer has been assumed to be isotropic. In this relationship, the bar sign over the letters indicates the following scaled and normalized quantities:

$$\bar{t} = t/\tau_o, \quad \bar{\ell} = \ell/\ell_o, \quad \bar{\kappa} = \kappa\, \ell_o, \quad \bar{w}_o = w_o/\ell_o, \quad \bar{L} = L/\ell_o \qquad (2)$$

$$\Delta \bar{g}_{vb} = \frac{\breve{g}_{vb} \ell_o}{g_\sigma^o}, \qquad \bar{\vartheta} = \vartheta/(E_o \ell_o), \qquad \chi = e|\hat{Z}| E_o \ell_o^2 /(\Omega_\sigma g_\sigma^o) \qquad (3)$$

The time and space variables $\{t, \ell\}$ have been scaled in the following fashion; first of all, $\hat{M}_\sigma$, an atomic mobility associated with mass flow at the surface layer, is defined and then a new time scale is introduced by $\tau_o = \ell_o^4 / \left( \Omega_\sigma^2 \hat{M}_\sigma g_\sigma \right)$, where $\ell_o$ is the arbitrary length scale, which is for the present simulation studies chosen as $\ell_o = 2w_o/3$, where $w_o$ is the half width of the interconnect specimen. $E_o$ denotes the electric field intensity directed along the specimen longitudinal axis, $e|\hat{Z}|$ is the effective charge, which may be given in terms of the atomic fractions, $x^i$, by $\hat{Z} = \sum_i x^i \hat{Z}^i$ for multi-component alloys. In the present study the generalized mobility, $\hat{M}_{vb}$ associated with interfacial displacement reaction taking place during the surface growth process (adsorption or desorption) is assumed to be independent from the orientation of the interfacial layer in crystalline solids. It is normalized with respect to the minimum value of the mobility of the surface diffusion denoted by $\hat{M}_\sigma$. They are given, respectively by: $\hat{M}_\sigma = \left( \tilde{D}_\sigma h_\sigma / \Omega_\sigma kT \right)$ and $\bar{M}_{vb} = \left( \hat{M}_{vb} \ell_o^2 \right) / \hat{M}_\sigma$. Here, $\bar{\Omega}_\sigma$ is the mean atomic volume of chemical species in the void surface layer. $\tilde{D}_\sigma$ is the isotropic part (i.e., the minimum value) of the surface diffusion coefficient. In the formulation of the problem, we have adapted the convention such that the positive direction of the motion is always towards the bulk material whether one deals with inner voids or outer surfaces or interfaces.

## 3. Results and Discussions

In our computer simulation studies, it is assumed that the sample was sandwiched between the top and bottom high resistance ($TiAl_3$, TiN, etc.) coatings, which constitute diffusion barrier layers together with the substrate. It is assumed here that only the one edge (sidewall) of the interconnect line is subjected to the surface drift-diffusion, and it is exposed to environment whose conductivity is neglected in this study. In certain cases [22], in which the upper surface of the unpassivated interconnect becomes predominant path for the drift-diffusion; the results of our computer simulation may still be applied by modifying the line width parameter, $w_o$, with the line thickness denoted by $h_o$.

Altogether 96 different combinations of the surface textures, drift-diffusion anisotropy coefficients, and EW intensity parameters are subjected to this simulation work on the hillocks, and only a few representative ones are reported in this paper. In all these cases, the direction of the EW is chosen along the longitudinal axis of the interconnect line to sweep the unperturbed sidewalls and upper and lower surfaces of the test piece. We have employed mainly four different normalized EW intensity parameters, ($\chi = 5, 10, 25$ and $50$) in the electromigration dominating regime (EMDR), which covers from the moderate up to the high current densities ($j = 10^9 - 10^{12}$ A/m$^2$) that are mostly utilized in normal and accelerated laboratory test studies. As an initial surface topology, a Gaussian shape hillock is introduced on top of the otherwise perfectly smooth flat edge surface. We have also adapted large aspect ratios, $\beta = 20 - 30$ with reference to the halve-width of the hillock present at the surface, and utilized quasi-infinite boundary conditions such as that the surface curvature and their higher derivates are all equal to zero at the anode and cathode edges. The constant and uniform electric field is applied to the specimen as a particular solution (initial data), which assures that there is a steady flow of atomic species from the cathode end to the anode edge of the specimen. In order to sustain a constant current condition all along the simulation experiment, the induced electric field (the complementary solution) due to surface disturbances and the EM induced internal voids are calculated by modified indirect boundary element method (MIBEM), using special Neumann boundary conditions which leaves the cathode and anode edges completely open for the constant current flow, and seals off the sidewall (edges) surfaces, and the EM induced internal voids from any current penetration.

The morphology of an initially perfectly flat surface having a perturbation in the shape of a Gaussian edge-hillock is demonstrated in Fig. 1, where the positive direction of electric field is from the left (anode) to the right (cathode). The scaled interconnect width is denoted as $\bar{w}$ and the void depth and the specimen length are given by $\bar{a}$ and $\bar{L}$, respectively. These are all scaled with respect to the arbitrary length denoted by $\ell_o$. Three different crystal planes, $\{110\}$, $\{100\}$ and $\{111\}$ for the surface of the single crystal thin films having a fcc crystal structure, are considered, which may be characterized by two, four and six fold symmetry zone axes, respectively.

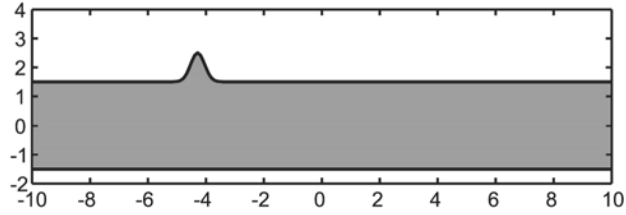

**Fig. 1.** Side view of a thin metallic single crystal film with a preexistent edge hillock just before the onset of the application of the electric field.

The macro behavior of the edge-hillock evolution kinetics in terms of the applied EW intensity and the degree of diffusion anisotropy is analyzed for each texture configuration in the following sub-sections, and the results are summarized in Tables 1, 2, and 3, for the two-fold, four-fold and six-fold symmetry planes. In general, the tilt angle between the direction of the electric field and the one of the principal axis of the diffusion dyadic is defined in the semi-closed interval.

In Fig. 2, the anisotropic diffusion coefficient and its first two higher derivatives are plotted against the tilt angle, for the four fold-symmetry planes of a fcc crystal structure, which will be used as a prototype later systematically in our discussions on the instability analysis of the finite surface perturbations. Since the present computer simulation experiments have been devoted for the EM dominating regime, namely $\chi > 1$, one can easily differentiate two distinct domains, (dissipative and regenerative) in terms of the tilt angle only, without going any complications with the capillary effects. There are two important exceptions, which are related to the transition stages taking place while passing from the stability to the instability region or vice versa, and they cannot be predicted by the linear instability analysis (LISA) theory [23] at high EW intensities especially in the case of edge-voids. These critical transitions in the tilt angles are corresponding to the extremum in the anisotropic diffusion coefficient denoted by $\phi = 0,$ and $\phi = \pi/m$, and plus their periodic extensions. According to the results of our computer simulations performed in the EM dominating regime, the stability and instability regimes for the finite amplitude perturbations may be defined by the following open intervals for the tilt angles: $(0<\phi<\pi/2m)$ and $(\pi/2m<\phi<\pi/m)$, and plus their periodic extensions represented by $\{n\pi/m\}$, where $n \leq m$ is a set of positive integer numbers, respectively.

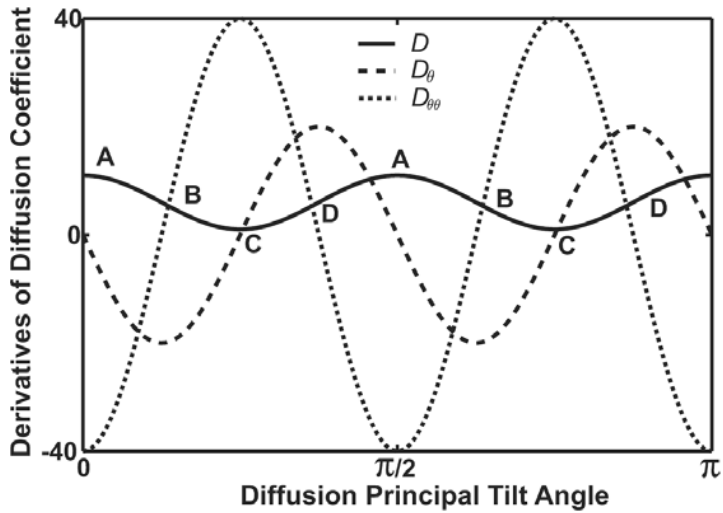

**Fig. 2.** Anisotropic diffusion constant and its higher order derivatives for the four-fold symmetry, {100}, plane of a fcc crystal. A: very unstable (the wedge shape fatal internal void); B: very stable; C-I: at the low EW intensities the surface disturbance decays; C-II: at the high EW intensities the formation of solitary wave train occurs; D: very unstable (oscillatory waves).

### 3.1. Two Fold Crystal Symmetry, {110} Planes in fcc:

Table 1 outlines the results of the extensive computer simulation experiments performed on the {110} oriented sidewalls of a fcc-type single crystal thin film, in a systematic fashion. In these simulations, various EW intensity parameters, tilt angles and diffusion coefficient anisotropy constants (*A=5* and *A=10*) are used. The interconnect line with zero degree tilt angle and having a Gaussian shape of edge-hillock shows two different evolution behaviors, depending upon the applied normalized EW intensity parameter denoted as $\chi$. At very low EW intensities $\chi \leq 5$, the hillock decays due time and completely disappears leaving behind almost perfect flat surface (Fig. 3).

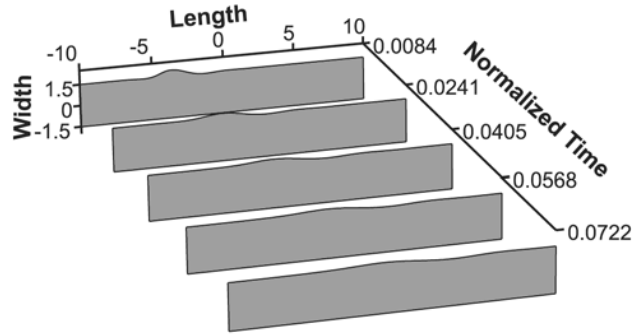

**Fig. 3.** Morphological evolution of a hillock, which shows steady decay, while drifting towards the cathode side with uniform velocity, and leaving behind almost flat surface. $\chi = 5;\ A = 5;\ \phi = 0^\circ;\ m = 1$.

The close inspection of the anisotropic diffusivity expression in Fig. 2 shows that for this tilt angle, the diffusion rate is maximum along the applied electric field, $D_\phi(0,\phi \to 0, m) = 0$ and $D_{\phi\phi}(0,\phi \to 0, m) < 0$, where $m = 1$. This behavior is in accord with the first order linear instability analysis (LISA) of the governing equation (Eq. 1) by Brush and Oren [23] for this tilt angle. LISA states that one should be in the stability regime since the contribution from the electromigration to the growth rate constant denoted by $\Gamma$ becomes equal to zero, and the remaining term due to the capillary force is negative (dissipation), and only contributes to the stability.

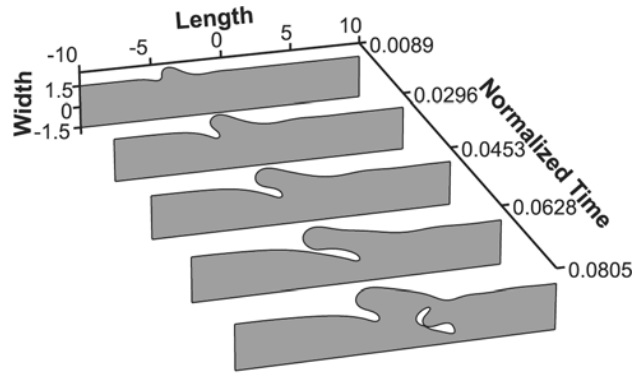

**Fig. 4.** The morphological evolution of a hillock under the moderate EW intensity. The hillock shows over hanging to leeward side, and pooping up internal void. The EM induced void shifting towards the cathode edge with a uniform velocity. $\chi = 25$; $A = 5$; $\phi = 0^\circ$; $m = 1$.

At moderate to high EW intensities $\chi \geq 10$, the situation is completely different (Fig. 4). The top of the hillock starts to over hangs on the leeward side. On the other hand, the lee side of the hillock becomes more and more protruding deep into the bulk region with almost 45° inclination towards to the cathode edge, and creating a bottleneck. Then, the intruding part of the hillock breaks up from the bottleneck portion and becomes an interior wedge void. This newly created inner void, the size of which is somehow inversely dependent on the applied EW intensity, drifts towards the cathode end. The remaining part of the hillock has still very long intrusion, which may act as a source for the multiple inner void generations by breaking-up. The normalized drift velocity of the EM induced internal void is plotted with respect to the electron wind intensity parameters $\chi$, for various values of the anisotropy parameter in Fig. 5.

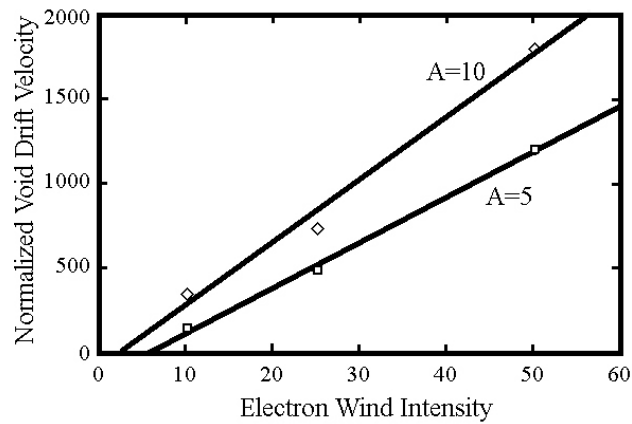

**Fig. 5.** Electromigration induced internal void drift velocity versus the electron wind intensity parameter for two different diffusion anisotropy constants; $A = 5$ and $A = 10$.

This figure clearly shows the linear connection between the velocity and the EW intensity, which may be represented by the following formula after the renormalization procedures:

$$V = \frac{Dh}{kT}\left\{ \alpha \frac{3e|Z|\rho J_o}{w_o} - \beta \frac{9\Omega g_\sigma}{w_o^2}\right\} \quad (4)$$

where, $\alpha \leftarrow [27;37]$ and $\beta \leftarrow [143;85]$ are constants obtained by the linear regression analysis of the data represented in Fig. 5 for the two anisotropy constants given by A=5 and A=10, respectively. $\rho$ is the specific resistivity, $g_\sigma$ is the surface specific Gibbs free energy, and $J_o$ is the applied current density. This expression also gives the threshold value of the applied electric field below which no inner EM induced void can form, namely: $E_o^{Thrs.} = \left\{3\Omega g_\sigma / \left(e|Z|w_o\right)\right\}$, and in term of the EW intensity parameter, which may be given by $\chi^{Thrs.} = \beta/\alpha \cong [2.30; 5.30]$ depends on the anisotropy constants, respectively.

We also investigated the tilt angle 45° in our simulation studies, and the typical results obtained for the two extreme EW intensities are illustrated in Figs. 6 and 7.

There are two distinct regimes for the same tilt angle, depending upon the applied electrostatic potential. For low and moderate values of EW intensity parameters $\chi = 5-25$, the edge-hillock slowly broadens without changing its general form and finally disappears with no trace behind (Fig. 6). The decay time depends on the electric field intensity inversely and the diffusion anisotropy linearly as can be seen in Table 1.

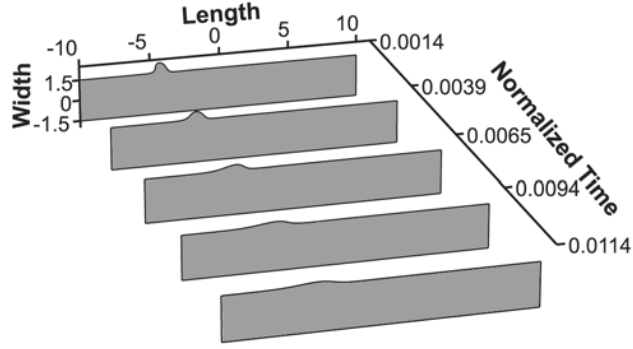

**Fig. 6.** Morphological evolution of a hillock at low EW intensities. The hillock shows decay in time, and leaving behind a perfect flat surface. $\chi = 5$; $A = 5$; $\phi = 45^\text{o}$; $m = 1$.

For the high EW intensities $\chi \geq 30$, a very interesting topological evolution occurs, which is first time observed in our computer simulation studies. Namely, the edge hillock starts to bend over the leeside with certain degree of extrusion due to very high EW intensity (normalized with respect to the capillary forces). Subsequently, this extruded part breaks up from the bottleneck portion, and becomes an autonomous entity of round shape (Fig. 7). To study the further behavior of this broken up piece of metal, we allow it to be exposed to the applied electric field utilizing the fact that underlayer can act as shunt. The result as shown in Fig. 7 indicates that the broken piece takes the perfect circular shape and slowly drags towards the anode side of the interconnect line. The surface after the ejection of this piece of material becomes almost flat without leaving any trace of the surface disturbances.

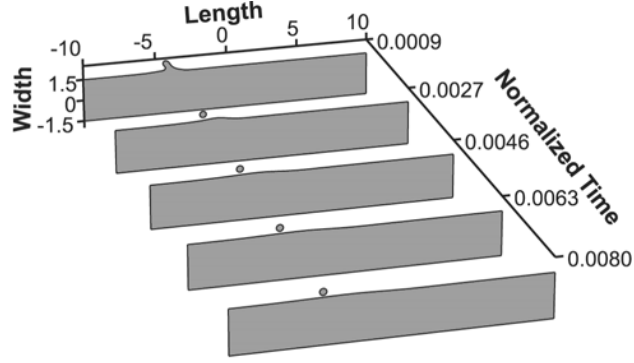

**Fig. 7.** Morphological evolution of a hillock under the high EW intensity. The hillock intrudes towards the leeward side and forming a bottleneck. The process ends up by a detachment of a piece of metal that may drift towards the anode if the underlayer acts as a shunt for the applied electric field. $\chi = 50$; $A = 5$; $\phi = 45^{o}$; $m = 1$.

The 90° tilting shows three distinct morphological evolutions depending on the normalized EW intensity. At low EW intensities $\chi \leq 5$, the edge-hillock transform into the kink shape surface void having sharp edge at the lee side, and drifts steadily towards the cathode end, with an almost constant velocity (Fig. 8a). At moderate EW intensities, $10 \leq \chi \leq 25$ there is a remarkable topological change on the hillock by assuming a thorn shape, and pointing towards the anode edge (Fig. 8b). This pointed throng, at very high EW intensities $\chi \geq 50$, becomes intruded to the lee side, and finally breaks up from the bottleneck region, and generates a tiny internal void (Fig. 8c). It seems that this process may repeat many times, since the leeward side of the edge void drifts much faster than the windward side. These both regions are combined together at the top by trapping an empty space, which is nothing but a formation of an inner void. These newly formed inner voids migrate towards to cathode end. Strangely enough, this tiny void doesn't follow a straight path but rather takes a route, which closely traces the contour line of the edge-void. This is due to the strong current crowding taking place between the inner void and the EM induced edge-void because of the extremely high electron current densities plus the very close proximities between these two defects.

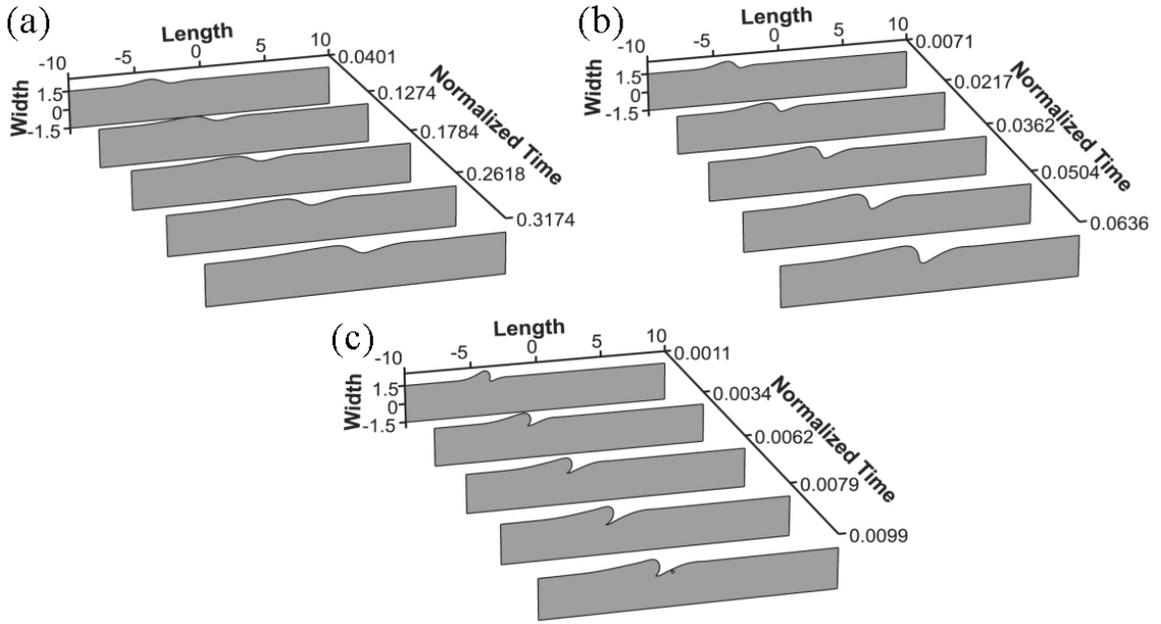

**Fig. 8.** The morphological evolution of a hillock under the moderate EW intensity. The hillock transform into thorn shape pointing towards the cathode edge, and moves with uniform velocity after stabilizing its amplitude (*the solitary wave*). $\chi = 5$ (a); 25 (b); 50 (c); $A = 5$; $\phi = 90^\circ$; $m = 1$.

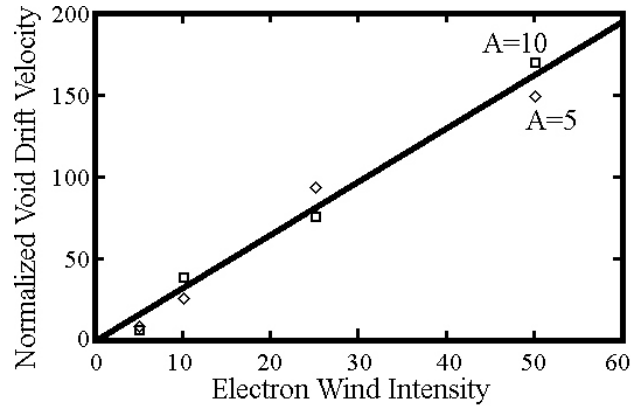

**Fig. 9.** The propagation velocity of the solitary wave with respect to the EW intensity shows a linear relationship, $v = 3.25\chi$, which is not affected very much by the values of the anisotropy constants used in this work; $A = 5$ and $A = 10$.

In Fig. 9, the propagation velocities of the solitary surface waves for various electron wind intensity parameters are plotted for two different diffusion anisotropy constants; A=5 and A=10. This figure clearly shows that there is a linear connection between velocity and the EW intensity parameter, which is not affected very much from the morphological variations in the shape of the surface waves as well as from the changes in the anisotropy constants used in this paper.

The tilt angle 135° always represents extreme instability and the formation of oscillatory and spreading wave packages on the lee as well as on the windward sides of the edge-hillock. The creation of the oscillatory waves doesn't cause any reduction of the strength of the initial hillock. Eventually for the moderate and high EW intensities, the windward edge of the hillock starts to intrude to the interior of the bulk region. This follows up by necking and ejection of an interior void as may be seen in Fig. 10.

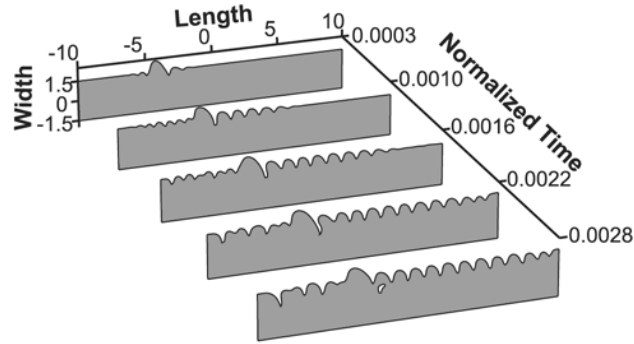

**Fig. 10.** The morphological evolution of an edge-hillock under the severe EW intensity. The shape of the initial hillock is almost stabilized by ejecting tiny internal void, while oscillatory waves are generating and spreading on the lee as well as on the windward directions. $\chi = 25$; $A = 5$; $\phi = 135°$; $m = 1$.

### 3.2. Four Fold Crystal Symmetry, $\{100\}$ planes in fcc:

The four fold crystal symmetry combined with the zero tilt angle may be very troublesome for interconnects. In all range of EW intensities, $\chi = 5-50$, studied in this work, the edge-hillock converts into a kink-like waveform. At moderate EW intensities $\chi = 10-25$, it starts to intrude towards the windward direction, with an inclination of $45°$ with respect to the longitudinal direction of the specimen, and than generates large internal void. This giant size void has very unusual form as illustrated in Fig. 11, namely; the wedge shape doublet. Definitely it causes a fatal break down of the interconnect line, when it hits to the opposite edge.

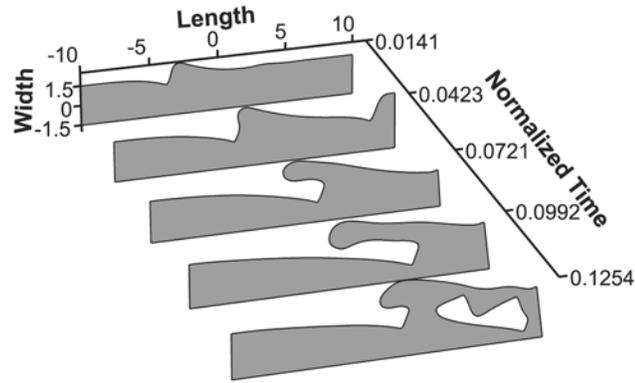

**Fig. 11.** Hillock morphological evolution into a regenerative intricate wave form, by first forming a positive kink, and than creating wedge shape EM induced inner void, by hanging over the leeward side and trapping some part of the vapor phase, at the moderate EW intensities. $\chi = 25; A = 5; \phi = 0^o; m = 2$.

In the case of strong anisotropy $A \geq 10$, instead of EM induced internal void formation, the multiplication of the sawtooth shape waves in front of the original edge-hillock takes place, with a constant increase in the wave amplitudes, and finally reaching the opposite sidewall and creating fatal failure of the interconnect line by breaching (Fig. 12).

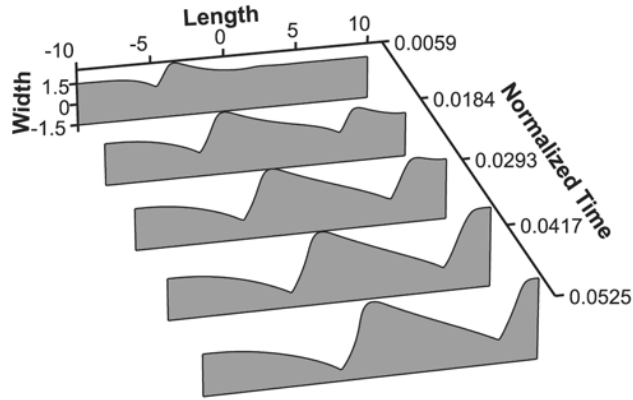

**Fig. 12.** Hillock morphological evolution into the train of sawtooth shape waves, with increasing amplitudes while traveling to the cathode end. Finally the waves reach to the opposite side of the interconnect and cause a fatal breaching. $\chi = 25;\ A = 10;\ \phi = 0^{o};\ m = 2$.

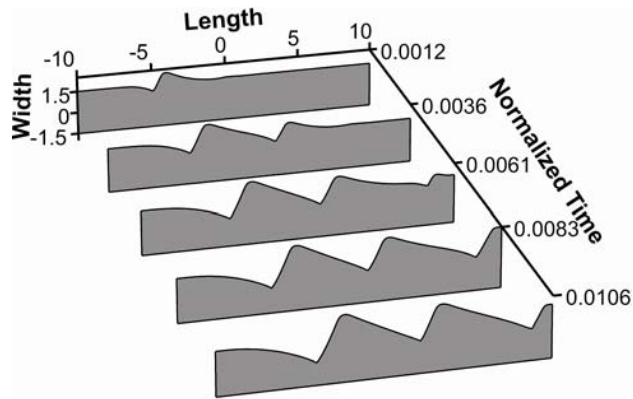

**Fig. 13.** Hillock morphological evolution into the train of sawtooth shape waves, with slowly increasing in amplitudes while traveling to the cathode end. Finally reaching the opposite side of the interconnect causing fatal breaching. $\chi = 50;\ A = 10;\ \phi = 0^{o};\ m = 2$.

In the case of very high EW intensities, $\chi \geq 50$, the amplitude of the individual sawtooth shape waves are stabilized in a short time by adapting almost perfect sawtooth shape, regardless of the strength of the diffusion anisotropy. These waves represent a *solitary wave* train and having much shorter wavelengths compared to the low and moderate EW intensities. This spreading and multiplying solitary wave package drifts towards the cathode end with a uniform velocity (Fig. 13).

As can be seen in Fig. 14, when the tilt angle becomes 30°, the edge-hillock at low and moderate EW intensities, $\chi = 5-25$, dies off gradually, with a minor modification in the shape such as a small bending towards the leeside. At high intensities, the Gaussian shape hillock transforms into a finger shape hillock bending towards the leeside. This bended finger shape hillock for the moderate diffusional anisotropy, $A = 5$, transforms into the step shape wave front, which shifts steadily towards the windward direction.

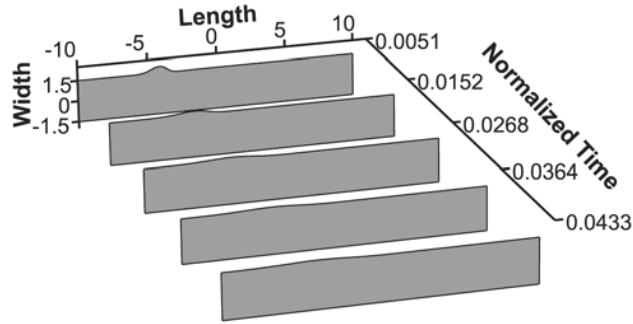

**Fig. 14.** Morphological evolution of an edge-hillock, which shows steady decay in amplitude, and finally leaving the perfect flat surface behind. $\chi = 10$; $A = 5$; $\phi = 30^\circ$; $m = 2$.

At high diffusional anisotropy, $A = 10$, the evolution behavior is completely different as illustrated in Fig. 15. Namely, it grows keeping its finger shape form with a slight bending to the leeward side and shifts rather fast towards the cathode edge.

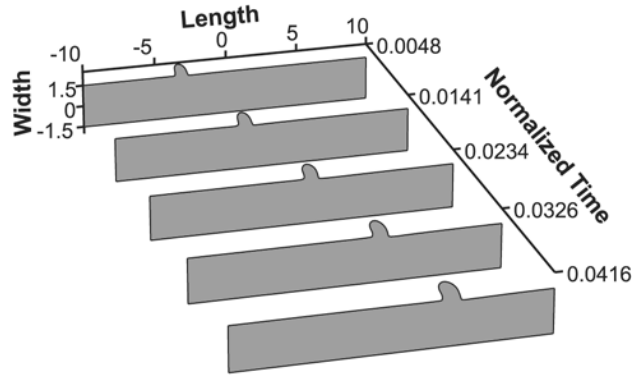

**Fig. 15.** Morphological evolution of a hillock to the finger shape extrusion (*solitary wave*) at very high EW intensity, which is subjected to the high degrees of diffusional anisotropy. The drift takes place with uniform velocity towards the windward side with steady increase in the amplitude, and the meanwhile the rest of the surface keeps its almost perfect flat shape. $\chi = 50$; $A = 10$; $\phi = 30°$; $m = 2$.

The general evolution behavior of the edge-Hillock at 45° tilt angle is illustrated in Figs. 16 and 17, for low and high EW intensities, respectively. In both extreme cases, the first edge-hillock transforms into the sawtooth morphology, and shifts towards the cathode end.

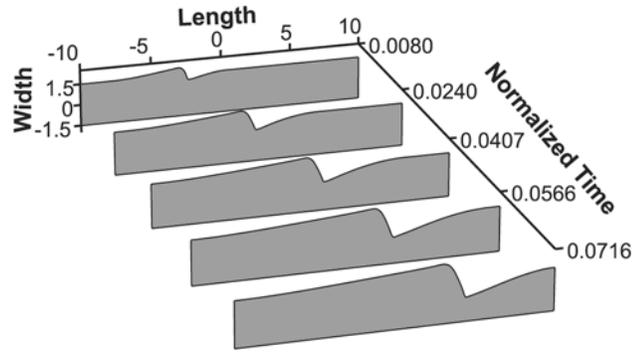

**Fig. 16.** Morphological evolution of hillock to sawtooth morphology at low EW intensity. The hillock drifts steadily towards the cathode edge. The amplitude of the hillock grows till it touches the opposite sidewall, and creating fatal breakdown in the interconnect thin film by breaching. $\chi = 5$; $A = 5$; $\phi = 45°$; $m = 2$.

In the case of high EW intensities, the multiplication in the number of sawtooth is very obvious while the whole wave package drifts steadily along the windward direction.

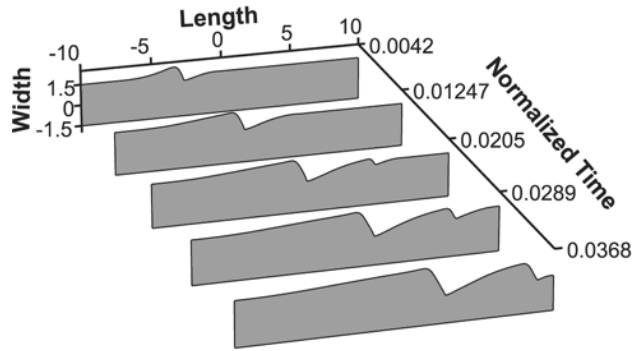

**Fig. 17.** Morphological evolution of a hillock to saw tooth morphology at very high EW intensity, with creation of new wavelets similar in shape during the steady drift towards the cathode edge. The amplitude of the original hillocks grows without limit till it may touch the opposite sidewall, and creating fatal breakdown in the interconnect thin film.

$\chi = 50$; $A = 5$; $\phi = 45^\circ$; $m = 2$.

When the tilt angle changes to 60°, the surface becomes very unstable, which almost corresponds to point D in Fig. 2, and the oscillatory waves on lee as well as on windward sides start to appear with increasing intensities. In later stages, the form of these waves transforms into the sawtooth shape having rather sharp windward front. A typical situation is illustrated in Fig. 18, where $\chi = 25$ and $A = 10$. The intensity of the electron winds doesn't affect the overall evolution behavior.

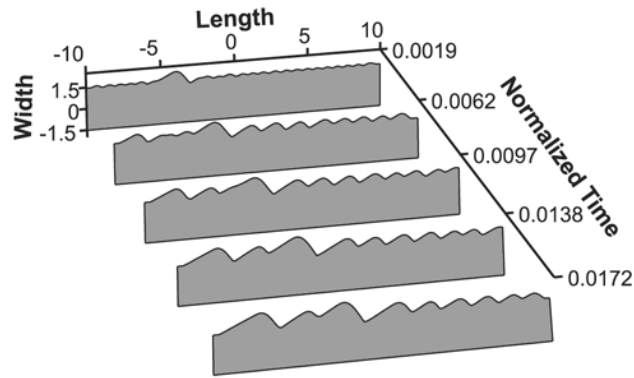

**Fig. 18.** Generation of the highly unstable oscillatory wave train out of an edge-hillock surface topography, which spreads and multiplies in both directions in growing amplitude. Meanwhile the original hillock (embedded in the background) moves towards the cathode with increasing in amplitude. $\chi = 25$; $A = 10$; $\phi = 60^\circ$; $m = 2$.

### 3.3. Six-Fold Crystal Symmetry, {111} Planes in fcc:

Finally, we have performed extensive simulation studies to investigate the morphological evolution behavior of the sidewalls of $\{111\}$ surfaces of thin metallic single crystal films having fcc structures, which are subjected to the Gaussian shape localized edge-hillocks, and oriented with various degrees of tilt angles with respect to the direction of the applied electric field. The surface drift-diffusion coefficient is chosen as anisotropic with the anisotropy coefficients in the range of $A = 5-10$, and the applied normalized EW intensity parameter spanning a large range of values, $\chi = 5-50$ but still staying in the EM dominating regime.

The specimen surfaces oriented with the zero tilt angle with respect to the wind direction show almost similar morphological evolutions, namely conversion of the Gaussian shape hillock into the kink shape wave front by inclining towards the windward side, and having constant increase in the amplitude while drifting towards the cathode direction. This behavior clearly indicates extreme instability, and causes the fatal break down in the interconnect as soon as it touches the opposite sidewall. At moderate EW intensities, $10 \leq \chi \leq 25$, the generation or the formation of the daughter kinks with ever increasing in number (multiplication) takes place in such a manner that their amplitudes are almost stabilized. This multiplication kinetics is extremely fast at very high EW intensities, $\chi \geq 50$, At the latter stage, the lee side of the converted mother hillock protrudes and finally hangs over itself creating a bottle neck. The large piece of material ruptures from this bottleneck, leaves the interconnect line, and drifts towards the anode end. This unusual behavior is illustrated in Fig. 19, where the underlayer is assumed to be acting as a shunt for the applied electrostatic field to the interconnect thin film to simulate this event numerically. Therefore, the broken away piece is still subjected to the applied EW after leaving the main body.

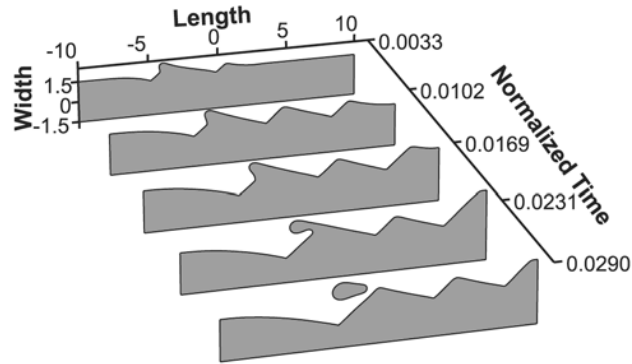

**Fig. 19.** Morphological evolution of edge-hillock into the kink-shape wave package, which is accompanied by the detachment of a metal piece That drifts in the direction of the anode edge with a uniform velocity as oppose to the kink motion. $\chi = 50;\ A = 5;\ \phi = 0^\circ;\ m = 3$.

After this rupture, the remaining part of the hillock becomes sawtooth in shape and continuous to drift to the cathode end. Meanwhile more and more subsidiary waves are generated on the windward side. The complete package, made up by sawtooth waves, drags towards the cathode end by constantly increasing the number of their members at the front (multiplication). However, this process breaks down when the lower edge of the preceding wave, which is nothing but the original hillock with different face, touches the opposite edge of the interconnect line.

For 15° tilt angle, the edge hillock transforms into step like wave, which slowly decays off while shifting towards the cathode side (Fig. 20). However, at moderate and high electron winds, $\chi \geq 25$, the step like wave starts to grow steadily, while drifting with a constant speed in the windward direction. Eventually, the lower edge of the wave front may touch the opposite edge of the interconnect line, and terminates the process completely (electrical shut down).

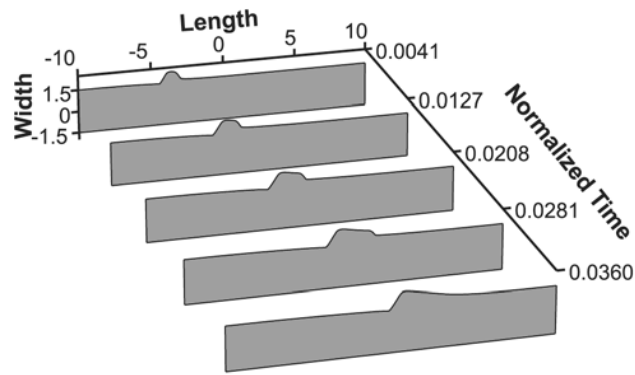

**Fig. 20.** At very high EW intensity, the morphological conversion of the hillock to the negative kink (*the Solitary wave train*) appears, which follows by the generation of the frontal wave package, which drifts towards the cathode edge with a uniform velocity. $\chi = 50;\ A = 5;\ \phi = 15^\circ;\ m = 3$.

The edge-hillock transforms into a negative kink-shape wave by lifting up the leeward side, when the tilt angle becomes 30°. This negative kink-wave drifts towards the cathode end with increasing intensity without showing any change in its form. In Fig. 21, the normalized velocity of the surface wave is plotted with respect to the electron wind intensity parameter for two different diffusion anisotropy constants.

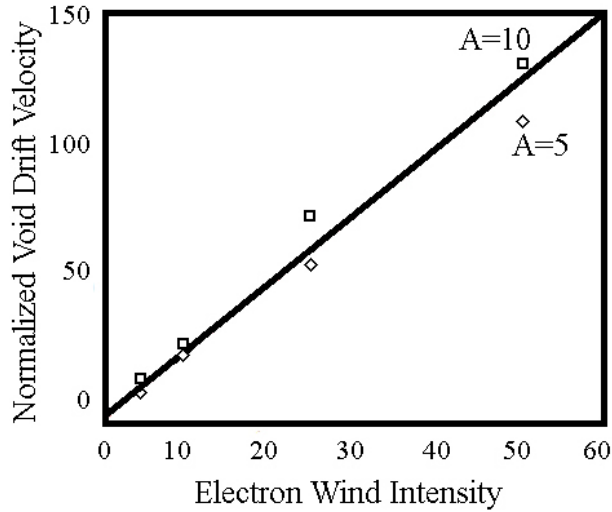

**Fig. 21.** The normalized velocity of train of solitary waves is plotted with respect to the electron wind intensity ($V = 2.5\chi$) for various diffusion anisotropy parameters. The tilt angle of the diffusion dyadic with respect to the electric field direction on the six fold symmetry plane is given by $\phi = 30^{\circ}$.

This figure again indicates a linear relationship, which is very similar to the one obtained for the two fold symmetry axis. The minor scattering in the data at high EW values is due to the anisotropy effect.

At high EW intensities such as $\chi \geq 50$, the multiplication takes place on the windward side. As may be seen in Fig. 22 that the overall picture looks like a chain of sawtooth waves shifting with uniform speed towards the cathode edge. This process may continue till the lower edge of the preceding wave touches the opposite side of the interconnect line, and breaking the electrical circuit.

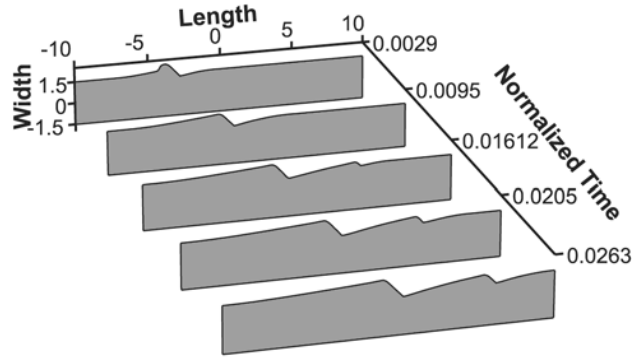

**Fig. 22.** Morphological evolution of a hillock into the negative kink-wave (*the solitary waves*), which generated frontal wave train with constant multiplications while moving with a uniform velocity towards the windward side. $\chi = 50$; $A = 5$; $\phi = 30^\circ$; $m = 3$.

We have also investigated a single crystal (fcc) interconnect line having $\{111\}$ surface structure with 45° tilt angle. The Gaussian shape hillock on this plane regardless of the intensity of the EW shows very unstable behavior. Immediately after it is exposed to the EW, the oscillatory waves are produced in both directions. As illustrated in Fig. 23, these waves grow, while they are drifting towards the cathode end. The anode end shows very much depletion at very high EW intensities due to the limited size of the interconnect length adopted in the present simulations experiments to save the computation time.

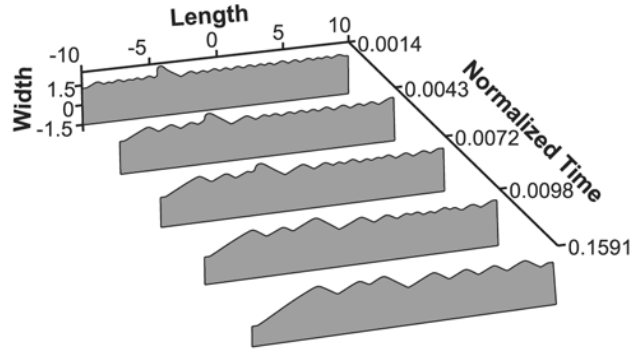

**Fig. 23.** Morphological evolution of hillock at very high EW intensity, and the generation of the frontal oscillatory wave train, which moves with a uniform velocity to the cathode edge. That is accompanied by the formation of a ditch on the anode edge, which may cause fatal break down in the interconnect in later stage. $\chi = 50;\ A = 5;\ \phi = 45^\circ;\ m = 3$.

## 4. Conclusions

These computer simulation experiments show that the degree of symmetry of the surface, represented by the fold number in the anisotropic crystal structure, is an extremely important factor in the determination of the morphological evolution of the sidewalls as well as the lifetime of thin film metallic single crystal interconnect lines. In addition to this prime factor, the orientation of the principal axis of the diffusion dyadic with respect to the direction of the electric field, which is represented by the tilt angle in our formulation, also plays a significant role in the final developments in the sidewall surface topology and even in the creation of the electromigration induced internal voids, which eventually cause the fatal break down in the electrical connection either hitting the opposite sidewall (breaching) or accumulating at the cathode end and resulting detrimental voiding at the contact area.

Since the present computer simulation experiments have been devoted for the electromigration dominating regime, namely $\chi > 1$, one can easily differentiate two distinct domains, (dissipative and regenerative) in terms of the tilt angle only, without going into any complications with the capillary effects. There are two important exceptions to this statement, which are related to the transitions taking place at the borderlines while passing from one regime to another. These critical transitions in the tilt angle are corresponding to the extremum in the anisotropic diffusion coefficient denoted by $\phi = 0$, and $\phi = \pi/m$, and plus their periodic extensions. The stability and instability regimes may be defined by specifying certain bounded but open sub-intervals of the tilt angle: ($0 < \phi < \pi/2m$) and ($\pi/2m < \phi < \pi/m$), and plus their periodic extensions represented by $\{n\pi/m\}$, where $n \leq m$ is a set of positive integer numbers, respectively.

In the interconnect metallic lines having bamboo or single crystal thin films, the most important failure mechanisms are closely associated with either the electromigration induced internal wedge shape voids generated by the surface instabilities caused by the anisotropy in the surface diffusivity and/or the cathode voiding by the transport of material through the surface (sidewalls) drift-diffusion driven by the electromigration forces. In both cases the surface drift-diffusion, which takes place at the interfacial layers or free surfaces, plays the predominant kinetic parameter for the determination of the lifetime. This parameter has to be controlled by some novel physicochemical processes such as alloying or utilizing certain liner or capping materials to reduce the atomic mobilities, increasing the interfacial Gibbs free energies, and/or decreasing the effective charge of the mobile species affected by the EM forces.

Finally, the selection of the most proper micro-texture with respect to the direction of the applied current flow to reduce the adverse effects of the diffusion anisotropy is crucial. It seems that the best choice for the device reliability considerations, which may be very difficult to achieve in practice by the epitaxial growth processes, would be the set of $\{110\}$ planes as a surface having about $45^o$ tilt angle, which has almost absolute sidewall surface stability even at very high current densities, and operating temperatures. The worst texture configuration would be, as far as the edge-wise breaching is concerned, the surface of the single crystal thin film which has six fold symmetry planes in fcc structures, namely $\{111\}$, which shows no counter acting dissipation mode regardless the orientation of the electric field with respect to the diffusion dyadic (tilt angle). In general, one may speculate that the choice of the high rotational symmetry planes for the upper surface of a thin single crystal film would be troublesome for any crystal structure that shows diffusional anisotropy with respect to those planes having the surface normal as a zone axis, as far as the edge-wise catastrophic breaching under EM forces is concerned.

**Acknowledgements**

The authors wish to thank Professor William D. Nix of Stanford University for his very valuable suggestions and motivations at the very beginning of this extensive research program at METU started in 1997. This research program is partially supported by the Turkish Scientific and Technical Research Council (TUBITAK) through the Project: No: 104M399.

# LIST OF TABLES

**Table 1.** The edge void morphological evolutions on the sidewalls of the {110} surface of a fcc single crystal thin metallic film, for various tilt angles, diffusion anisotropy constants, and the EW intensity parameters. The fold number is 2m=2.

| $\theta$ | | **0** | **45** | **90** | **135** |
|---|---|---|---|---|---|
| $\chi$ | **A** | 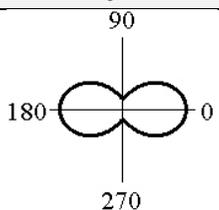 | 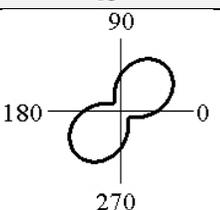 | 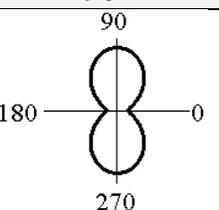 | 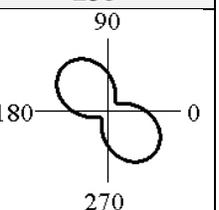 |
| 5 | 5 | D<br>$t_d = 0.03404$ | D<br>$t_d = 0.02509$ | S<br>(With growth) | OS<br>$t_s = 0.0221$ |
|  | 10 | D<br>$t_d = 0.0204$ | D<br>$t_d = 0.01635$ | S<br>(With growth) | OS<br>$t_s = 0.0103$ |
| 10 | 5 | V<br>$t_v = 0.06384$ | D<br>$t_d = 0.01627$ | S<br>(With growth) | OS<br>$t_s = 0.0165$ |
|  | 10 | V<br>$t_v = 0.03606$ | D<br>$t_d = 0.007115$ | S<br>(With growth) | OS<br>$t_s = 0.0034$ |
| 25 | 5 | V<br>$t_v = 0.00953$ | D<br>$t_d = 0.01369$ | S<br>(With growth) | OS<br>$t_s = 0.0045$ |
|  | 10 | V<br>$t_v = 0.006703$ | D<br>$t_d = 0.004735$ | S<br>(With growth) | OS<br>$t_s = 0.0016$ |
| 50 | 5 | V<br>$t_v = 0.0028$ | *Detachment of piece of metal*<br>$t=0.002095$ | V<br>$t_v = 0.01125$ | OS<br>$t_s = 0.001$<br>V$t_v = 0.00321$ |
|  | 10 | V<br>$t_v = 0.00181$ | *Detachment of piece of metal*<br>$t=0.0009407$ | V<br>$t_v = 0.00946$ | OS<br>$t_s = 0.0005$ |

D: decay, S: solitary wave, OS: oscillatory wave, V: internal void, F: rapture, and G: growth.

**Table 2.** The edge void morphological evolutions on the sidewalls of the {100} surface of a fcc single crystal thin metallic film for various tilt angles, diffusion anisotropy coefficients and the EW intensity parameters. The fold number is 2m=4.

| $\chi$ | $A$ | $\theta$ = 0 | 30 | 45 | 60 |
|---|---|---|---|---|---|
|  |  | 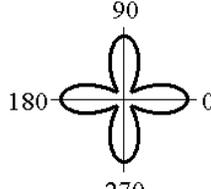 | 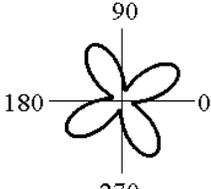 | 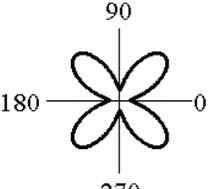 | 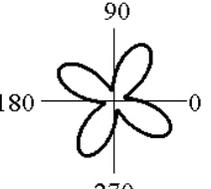 |
| 5 | 5 | K | D<br>$t_d = 0.03638$ | S / G | OS<br>$t_s = 0.01259$ |
|  | 10 | K | D<br>$t_d = 0.01725$ | S / G | OS<br>$t_s = 0.00635$ |
| 10 | 5 | K | D<br>$t_d = 0.03604$ | S / G | OS<br>$t_s = 0.004659$ |
|  | 10 | K | D<br>$t_d = 0.01894$ | S / G | OS<br>$t_s = 0.002054$ |
| 25 | 5 | V<br>$t_v = 0.11135$ | D<br>$t_d = 0.01711$ | S / G | OS<br>$t_s = 0.001345$ |
|  | 10 | F<br>$t_f = 0.0525$ | D<br>$t_d = 0.0175$ | S / G | OS<br>$t_s = 0.0006202$ |
| 50 | 5 | K / G | Hillock move to cath. with decay $t_d = 0.11818$ | Double edge void | OS<br>$t_s = 0.000358$ |
|  | 10 | K / G | Hillock move to cathode without decay | Double edge void | OS<br>$t_s = 0.000226$ |

D: decay, K: kink solitary wave, OS: oscillatory wave, V: internal void formation, and F: rapture.

**Table 3.** The edge void morphological evolutions on the sidewalls of the {111} surface of a fcc single crystal thin metallic film for various tilt angles, diffusion anisotropy coefficients, and the EW intensity parameters. The fold number is 2m=6.

| $\chi$ | | $\theta$ | 0 | 15 | 30 | 45 |
|---|---|---|---|---|---|---|
| | | A | 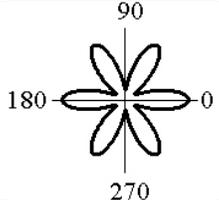 | 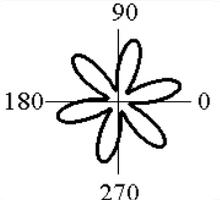 | 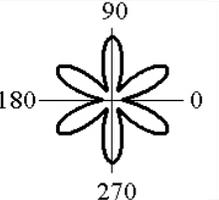 | 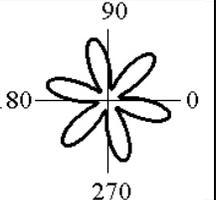 |
| 5 | 5 | | K / F $t_f = 1.112$ | D $t_d=0.0329$ | K / G | OS $t_s = 0.0094$ |
| | 10 | | K / F $t_f = 0.4152$ | D $t_d=0.03445$ | K / G | OS $t_s = 0.0030$ |
| 10 | 5 | | K | D $t_d=0.04625$ | K / G | OS $t_s = 0.0025$ |
| | 10 | | K | D $t_d=0.01556$ | K / G | OS $t_s = 0.0015$ |
| 25 | 5 | | K / F $t_f = 0.3105$ | D / ST | K / G | OS $t_s = 0.00056$ |
| | 10 | | K / F $t_f = 0.07425$ | ST / F $t_f = 0.1264$ | K / G | OS $t_s = 0.00028$ |
| 50 | 5 | | Detachment of piece of metal $t= 0.0285$ | V $t_d=0.002478$ | K / G | OS $t_s = 0.00021$ |
| | 10 | | K | D / ST | K / G | OS $t_s=0.000113$ |

K: kink shape wave, OS: oscillatory wave, ST: step shape wave, V: internal void formation, G: growth, and F: rapture.